\documentclass[aps,pra,twocolumn,nofootinbib,floatfix,10pt]{revtex4-2}
\pdfoutput=1
\usepackage[braket, qm]{qcircuit}
\usepackage{bbold}
\usepackage{appendix}
\usepackage{amsmath}
\usepackage{amssymb}
\usepackage{wasysym}
\usepackage{graphicx}
\usepackage{placeins}
\usepackage{color,soul}
\usepackage{siunitx}
\usepackage{dsfont}
\usepackage{xcolor}
\usepackage[english]{babel}
\usepackage{blindtext}
\usepackage[english,nomargin,inline,marginclue,draft]{fixme}
\pdfpageheight\paperheight
\pdfpagewidth\paperwidth

\usepackage[colorlinks,linkcolor=blue,anchorcolor=blue,citecolor=blue,urlcolor=blue]{hyperref}

\fxusetheme{colorsig}
\FXRegisterAuthor{cg}{acg}{CG}  
\FXRegisterAuthor{th}{ath}{\color{blue}TH}  
\FXRegisterAuthor{ib}{aib}{\color{red}IB} 
\FXRegisterAuthor{sh}{ash}{\color{cyan}SH} 
\FXRegisterAuthor{db}{adb}{\color{green}DB} 
\FXRegisterAuthor{ps}{aps}{PS}
\makeatletter
\renewcommand*\FXLayoutInline[3]{%
  {\@fxuseface{inline}\ignorespaces{\color{fx#1}[#3: #2]}}}
\makeatother

\long\def\symbolfootnote[#1]#2{\begingroup%
\def\thefootnote{\fnsymbol{footnote}}\footnotetext[#1]{#2}\endgroup}

\def\nobreakbefore{%
  \relax\ifvmode\else
    \ifhmode
      \ifdim\lastskip > 0pt\relax
        \unskip\nobreakspace
      \else 
        \nobreakspace
      \fi
    \fi
  \fi
}
\let\oldcite\cite
\renewcommand\cite{\nobreakbefore\oldcite}

\begin{document}
\title{Diatomic and Polyatomic Heteronuclear Ultralong-Range Rydberg Molecules}

\author{Qing Li$^{1,2}$}
\author{Shi-Yao Shao$^{1,2}$}
\author{Li-Hua Zhang$^{1,2}$}
\author{Bang Liu$^{1,2}$}
\author{Zheng-Yuan Zhang$^{1,2}$}
\author{Jun Zhang$^{1,2}$}
\author{Qi-Feng Wang$^{1,2}$}
\author{Han-Chao Chen$^{1,2}$}
\author{Yu Ma$^{1,2}$}
\author{Tian-Yu Han$^{1,2}$}
\author{Dong-Sheng Ding$^{1,2,\textcolor{blue}{\star}}$}
\author{Bao-Sen Shi$^{1,2}$}

\affiliation{$^1$Key Laboratory of Quantum Information, University of Science and Technology of China, Hefei, Anhui 230026, China.}
\affiliation{$^2$Synergetic Innovation Center of Quantum Information and Quantum Physics, University of Science and Technology of China, Hefei, Anhui 230026, China.}

\date{\today}

\symbolfootnote[1]{Corresponding author: dds@ustc.edu.cn}

\begin{abstract}
Ultra-long-range Rydberg molecules (ULRMs) have attracted significant interest due to their unique electronic properties and potential applications in quantum technologies. We theoretically investigate the formation and characteristics of heteronuclear ULRMs, focusing on Rb-Cs systems. We explore the vibrational energy levels of heteronuclear \(nD\) ULRMs and compare them with homonuclear counterparts. We also predict the formation of polyatomic heteronuclear ULRMs, discussing how the binding energy and spectral features evolve as the number of ground-state atoms increases. Our theoretical predictions are presented in terms of molecular spectra and provide insight into the formation dynamics of these systems. The study further explores the potential applications of heteronuclear ULRMs in quantum information processing, quantum simulation, and precision measurements, offering new avenues for future research in many-body physics and quantum technologies.
\end{abstract}

\maketitle

\section{Introduction}

Since the proposal by Greene et al. in 2000 \cite{greene2000creation}, ultralong-range Rydberg molecules (ULRMs) have become a significant focus of research in quantum and molecular physics. Unlike conventional molecules that are bound by covalent or ionic interactions, ULRMs are formed through a unique bonding mechanism driven by the scattering interaction between Rydberg electrons and ground-state atoms. This novel bonding mechanism allows ULRMs to achieve molecular sizes comparable to the Rydberg atoms that constitute them, typically on the order of \(100\) \(nm\). Such large molecular sizes and the distinctive bonding mechanism have sparked considerable interest in ULRMs, as they are expected to provide new insights into molecular behaviors and quantum effects \cite{lombardi2006dynamical, greene2006experimental, bendkowsky2009observation}.

A prominent example of ULRMs is the ``trilobite" molecule, which forms when a Rydberg electron localizes a ground-state atom within its orbit. This localization results in large permanent electric dipole moments (PEDMs) on the order of thousands of Debye in homonuclear ULRMs. The large size of these molecules, combined with the fundamental breaking of parity symmetry within the system, gives rise to these sizable dipole moments \cite{greene2000creation}. The existence of these large PEDMs enables experimental studies into their interaction with external electromagnetic fields\cite{kurz2013electrically, matzkin2001rydberg, wang2005alternative, lesanovsky2006ultra}. The dipole moments can be probed using electric fields, with spectral line broadening serving as an observable signature of their dipole interactions \cite{booth2015production, li2011homonuclear, althon2023exploring, niederprum2016observation}. Recent advancements also show that the spatial orientation of these molecules can be manipulated via external fields, enabling precise control of their alignment \cite{hummel2019alignment, gonzalez2015rotational, krupp2014alignment}. In addition, polyatomic ULRMs have been proposed and experimentally studied in recent years. Polyatomic ULRMs are formed when multiple ground-state atoms are captured by the large orbit of a Rydberg atom. Such interactions can lead to the formation of complex polyatomic molecules, which have been successfully demonstrated experimentally\cite{rittenhouse2010ultracold, rittenhouse2011ultralong, gonzalez2015rotational, gaj2014molecular,fey2019effective,luukko2017polyatomic}. The lifetimes of ULRMs have been studied, with results indicating that their lifetimes are primarily determined by the lifetime of the constituent Rydberg atoms \cite{butscher2011lifetimes, junginger2013quantum, camargo2016lifetimes}. 

The study of ULRMs, which offers a broader range of tunable molecular properties compared to homonuclear ULRMs, has recently gained significant attention. These properties include large dipole moments, varied binding energies, and more complex interactions with external fields. Such flexibility makes heteronuclear ULRMs an exciting area for both theoretical research and experimental exploration \cite{whalen2020heteronuclear, eiles2018twocomponent, peper2021heteronuclear, wojciechowska2024ultralong}. 

In this paper, we focus on investigating the formation and properties of diatomic and polyatomic heteronuclear ULRMs. By calculating potential energy curves (PECs), vibrational energy levels, electronic probability densities and PEDMs, we aim to gain insights into the behavior and interactions of diatomic heteronuclear ULRMs. Unlike homonuclear systems, heteronuclear systems provide a unique opportunity to manipulate atomic composition, allowing for more precise control over molecular characteristics and behavior. This ability to fine-tune molecular properties is especially valuable for applications in study of pair-correlation functions in atomic mixtures\cite{eiles2018twocomponent, whalen2020heteronuclear}, creation ultracold ion-pair systems with variable mass ratio and the dynamics of complex many-body heteronuclear systems\cite{peper2020formation, giannakeas2020dressed}.  In addition, this paper is the first exploration of the formation of polyatomic heteronuclear ULRMs in high-density Cs and Rb cold atomic gases and the theoretical calculation of their experimental spectra. The advantage of polyatomic heteronuclear ULRMs is their increased complexity, which provides a richer platform for investigating multi-body interactions and the emergence of novel quantum phenomena in strongly correlated systems.

\section{Calculation}
\subsection{Hamiltonian}

ULRMs consist of a Rydberg atom and one or more ground-state atoms, where the ground-state atom is bound to the electron orbit due to low-energy scattering interactions with the electron\cite{greene2000creation}. Firstly, we consider the case of a single ground-state atom, i.e., diatomic ULRMs. Following the Born-Oppenheimer approximation, if we take the atomic nucleus as the origin of coordinates, with the Rydberg electron's coordinate denoted as \(\vec{r}\) and the ground-state atom's coordinate denoted as \(\vec{R}\), the molecular Hamiltonian can be expressed as:
\[
H = H_0(\vec{r}) + V_{ea}(\vec{R}-\vec{r}) +V_{aa}(\vec{R}) + H_{vib},
\]
here, \( H_0(\vec{r}) \) represents the Hamiltonian of the isolated Rydberg atom, \( V_{ea}(\vec{R} - \vec{r}) \) represents the scattering interaction between the Rydberg electron and the ground-state atom, \( V_{aa}(\vec{R}) \) represents the interaction between the atomic core and the ground-state atom, and \( H_{vib} \) represents the vibrational energy of the molecule. The interaction \( V_{aa}(\vec{R}) \) between the atomic core and the ground-state atom is proportional to \( 1/\vec{R}^4 \). In a Rydberg system, \( \vec{R} \) can often be on the order of thousands of atomic units, so this term is reasonably neglected. Additionally, the rotational energy of the system, the electron spin, and the hyperfine structure are all neglected. Therefore, the total hamiltonian of ULRMs can be considered as consisting of three parts, which can be written as:
\begin{equation}
H = H_0(\vec{r}) + V_{ea}(\vec{R} - \vec{r}) + H_{vib},
\label{eq:H_total}
\end{equation}
here, the eigenenergies of the first two terms, as functions of \( \left|\vec{R}\right| \), represent the PECs under the Born-Oppenheimer approximation, which is one of the key focuses in the study of ULRMs. The last term represents the vibrational energy of the ULRMs, which is related to molecular spectra which is the most direct method of observing ULRMs. In this paper, unless otherwise noted, all expressions and calculations are made in atomic units.

In Eq. (\ref{eq:H_total}), the first term represents the Hamiltonian of the isolated Rydberg atom, i.e. the kinetic and potential energy of the Rydberg electron in the field of the atomic core. For alkali metal atoms, the eigenstates have energies given by the Rydberg formula,
\(E_{nl} = -{1}/{2(n - \delta_l)^2}\)
and wave functions \( \phi_{nl}(\vec{r}) \), where \( n \) and \( l \) are the principal quantum number and the orbital angular momentum quantum number, respectively. In the Rydberg state, \( \delta_l \) depends almost exclusively on \( l \), and the specific values can be found in \cite{eiles2019trilobites}. For the second term, following Fermi's approach, this interaction is expressed as a zero-range pseudopotential, and the \(s\)-wave and \(p\)-wave representations are given by\cite{fermi1934fermi, omont1977theory, greene2000creation, hamilton2002shape}:
\begin{equation}
V_{ea}(\vec{R} - \vec{r}) = 2\pi a_s(k)\delta(\vec{R} - \vec{r}) - 6\pi a_p^3(k) \delta(\vec{R} - \vec{r}) \overleftarrow{\nabla} \cdot \overrightarrow{\nabla},
\label{eq:V}
\end{equation}
here, \(a_s(k)\) and \(a_p^3(k)\) represent the low-energy scattering length and volume, respectively, which play a crucial role in determining the interaction strength. These quantities are related to the phase shifts \( \delta_s(k) \) and \( \delta_p(k) \) of a free electron with wave number \( k \) scattering off the ground-state atom (the Rydberg electron can be viewed as a free electron in this context), via the relations:
\[
a_s(k) = -\frac{\tan(\delta_s(k))}{k}, \quad a_p^3(k) = -\frac{\tan(\delta_p(k))}{k^3},
\]
the phase shifts \(\delta_s(k)\) and \(\delta_p(k)\) can be calculated exactly \cite{eiles2018formation, khuskivadze2002adiabatic, bahrim20013se, bahrim2001negative}. Specifically, the wave number dependence of the \(s\)-wave scattering length \(a_s(k)\) can be briefly described using the effective range formula \cite{o1962low}:
\[
a_s(k) \approx a_0 + \frac{\alpha \pi}{3} k,
\]
where \( a_0 \) is the zero-energy scattering length, and \( \alpha \) is the polarizability of the ground-state atom. For Rb(Cs) atom, the values of \( a_0 \) and \( \alpha \) used in this calculation are -18.5(-21.7) a.u. and 319.2(400.8) a.u., respectively\cite{gregoire2015measurements, bendkowsky2009observation, booth2015production}. The values of \( a_0 \) needs to be determined through precise experiments, and thus, experimental studies of ULRMs can also investigate the scattering between the electron and the ground-state atom. Additionally, the wavenumber of the Rydberg electron, \( k \), is expressed in semiclassical form as \(k(r) = \sqrt{2\left( E_{nl} + 1/r \right)}\), derived from energy conservation \cite{omont1977theory}.

\begin{figure}[h]
\centering
\includegraphics[width=1\linewidth]{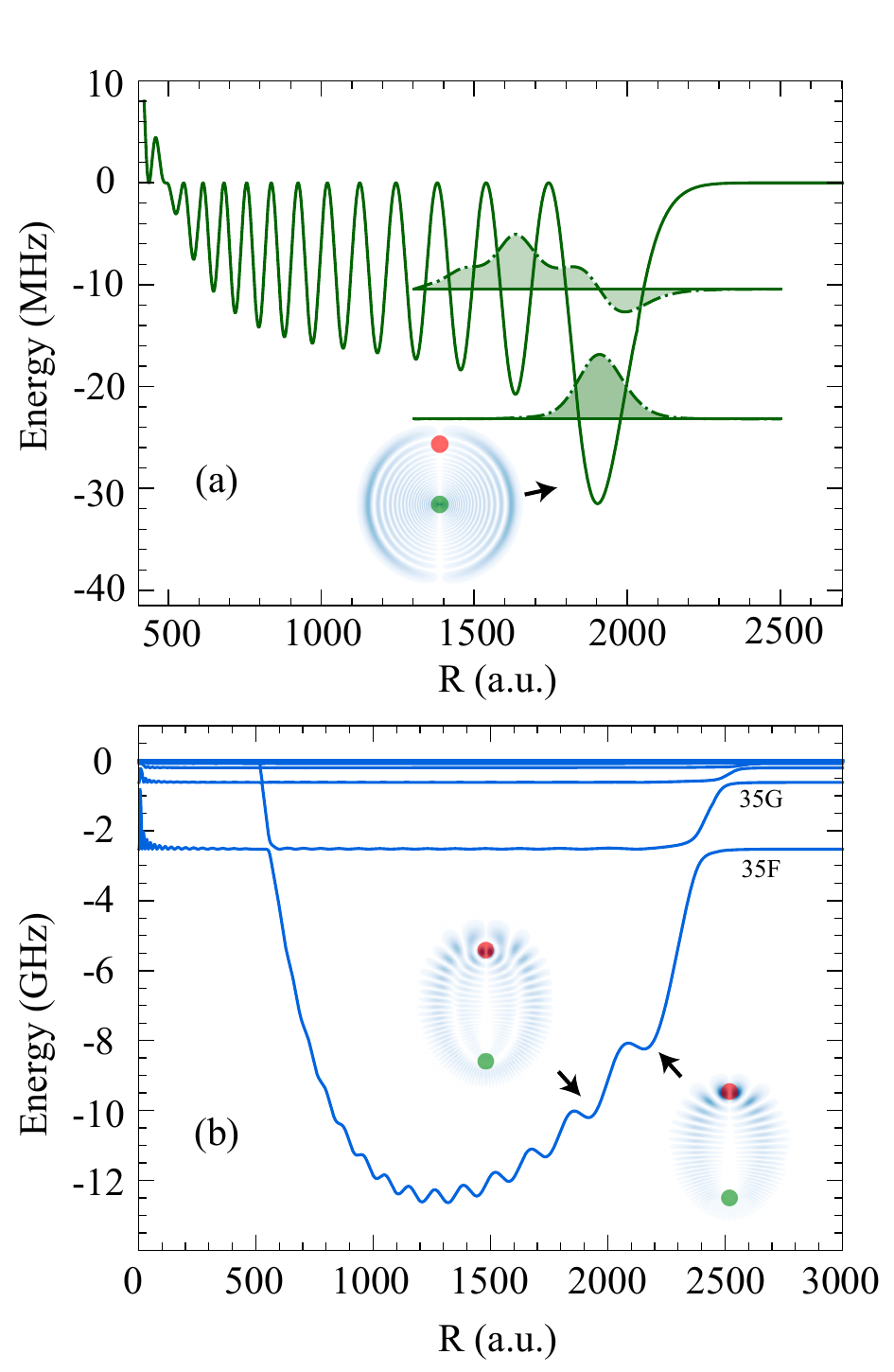}
\caption {
Rb(\(n=35\)) ULRMs are formed solely through \(s\)-wave scattering. (a) The green solid line shows the PECs of the Rb(\(35S\)) state, with the green dashed line representing the vibrational wave function. The horizontal lines near the dashed line represent the vibrational ground state and first excited state levels, shaded in light green. The shifts relative to the Rb(\(35S\)) Rydberg line are -23.18 MHz and -10.44 MHz, respectively, corresponding to the ground state and first excited state. The inset displays typical electronic probability density plots of the potential well at the arrow-marked positions in cylindrical coordinates for ULRMs: \( \rho |\psi(\rho, z, 0)|^2 \) and \( \rho |\psi(\rho, z, \pi)|^2 \). Here, the red (green) spheres indicate the positions of the ground-state atom (atomic core), and the \(z\)-axis is along the line connecting them. (b) The blue solid line shows the PECs for high-\( l \) Rb ULRMs, with the energy reference point being the \(n=35\) line of the hydrogen atom. The inset displays two typical electronic probability density plots at \( R = 1910 \, \text{a.u.} \) and \( R = 2150 \, \text{a.u.} \).
}
\label{Rb(35)简并和非简并微扰}
\end{figure}

\subsection{Only \(S\)-wave interaction}
In general, the low-\( l \) (\( l < 3 \)) states of alkali-metal atoms are isolated, while the high-\( l \) (\( l \geq 3 \)) states are nearly degenerate, and there is no degeneracy between low-\( l \) and high-\( l \) states. Therefore, to calculate the PECs associated with low-\( l \) electronic Rydberg states, non-degenerate perturbation theory can be applied. Following non-degenerate perturbation theory, the first term in Eq. (\ref{eq:H_total}) serves as the zeroth-order Hamiltonian, and the second term serves as the perturbation. Thus, \( E_{nl} \) and \( \phi_{nl}(\vec{r}) \) are regarded as the zeroth-order energies and zeroth-order wave functions. These energies and radial wave functions \( R_{nl} \) of these states can be obtained using the ARC-Alkali-Rydberg-Calculator package\cite{vsibalic2017arc}. The angular component can be represented as \( Y_{l} = \sqrt{{2l + 1}/{4\pi}} \), where \( l \) is the orbital angular momentum quantum number. In the case of only \(s\)-wave scattering, the first-order corrections to the energy and wave function can be expressed as\cite{sakurai1986modern}:
\[
V_{nl}(R) = \langle nl |2\pi a_s(k)\delta(\vec{R} - \vec{r})| nl \rangle,
\]
\[
\Phi_{nl}(R, \vec{r}) = \varphi_{nl}(\vec{r}) + \sum_{n'l'} \frac{\langle n'l' | 2\pi a_s(k)\delta(\vec{R} - \vec{r})| nl \rangle}{E_{nl} - E_{n'l'}} \varphi_{n'l'}(\vec{r}).
\]
For each \( R \), \( V_{nl}(R) \) has a value, and by plotting these values as a curve, we obtain the PECs for low-\( l \) ULRMs. The corresponding wave function density at each \( R \) can also be visualized. 

Additionally, the third term in Eq. (\ref{eq:H_total}) represents the molecular vibrational Hamiltonian. Based on the Born-Oppenheimer approximation, the molecular vibrational energy levels and vibrational wave functions can be computed. In the center of mass coordinate system for the Rydberg and ground-state atoms, the vibration of the ULRMs can be modeled as the motion of a particle with reduced mass \( \mu = (m_1 m_2)/(m_1 + m_2) \) in PECs, where \( m_1 \) is the mass of the Rydberg atom and \( m_2 \) is the mass of the ground-state atom. The dynamics of this reduced particle can be described by the Schrödinger equation\cite{sakurai1986modern}:
\[
\left[-\frac{\nabla_R^2}{2\mu} + V_{nl}(R)\right] \chi_v(R) = E_v \chi_v(R),
\]
where \( V_{nl}(R) \) represents the previously calculated PECs, and \( E_v \) and \( \chi_v(R) \) denote the \(v-th\) vibrational energy levels and vibrational wave functions, respectively. When performing calculations in atomic units, the electron mass is 1 a.u., and the masses of the Rb and Cs atoms are 158432 and 242282 a.u.. The PECs and vibrational wave functions of low-\( l \) ULRMs molecules are shown in Fig. \ref{Rb(35)简并和非简并微扰}(a). Under suitable experimental conditions, exciting the atoms to the Rydberg state allows for the observation of the molecular spectrum derived from this analysis \cite{bendkowsky2009observation, li2011homonuclear, tallant2012observation}. Because the energy levels of low-\( l \) states are relatively isolated, the electronic probability function of ULRMs is almost the same as that of the atomic state, as shown in the inset of Fig. \ref{Rb(35)简并和非简并微扰}(a). Furthermore, the calculations presented above reveal that, despite the relatively slow motion of the highly excited Rydberg electron, the Born–Oppenheimer approximation remains generally valid. This is because the energy spacing between adjacent Rydberg levels (on the order of GHz) is typically several orders of magnitude larger than the vibrational energy spacing of the nuclei (on the order of MHz).

For high-\( l \) states, because of their very small quantum defects, their energies are nearly degenerate. In this case, strong coupling between states arises, and the non-degenerate perturbation theory mentioned earlier is no longer applicable. We must use degenerate perturbation theory in this high-\( l \) subspace. In degenerate perturbation theory, we need to construct the Hamiltonian matrix, where the basis vectors chosen for the matrix are all the high-\( l \) states. The Hamiltonian matrix elements can be expressed as\cite{sakurai1986modern}:
\begin{equation}
H_{ij} = E_{ij} \delta_{ij} + [2\pi a_s R_i R_j] Y_i Y_j,
\label{eq:H}
\end{equation}
where \(i\) and \(j\) refer to the matrix elements corresponding to different \(l\) values. For each \(R\), a Hamiltonian matrix can be constructed and diagonalized to obtain eigenenergies and eigenvectors, with the number of these corresponding to the number of chosen states. Like low-\( l \) states, these eigenenergies as a function of \(R\) are the PECs of high-\( l \) ULRMs, as shown in Fig. \ref{Rb(35)简并和非简并微扰}(b). Moreover, once the eigenvectors are obtained, the Electronic probability density for the ULRMs can be expressed as a superposition of the wave functions of the basis vectors based on these eigenvectors\cite{sakurai1986modern}:
\begin{equation}
\Phi(\vec{r}, {R}) = \sum c_{l}({R}) \, \phi_{nl}(\vec{r}),
\label{叠加}
\end{equation}
in this equation, \(\Phi(\vec{r}, {R})\) denotes the electronic probability density function of ULRMs, \( c_{l}({R}) \) represents the eigenvectors, serving as the coefficients for the superposition, and \( \phi_{nl}(\vec{r}) \) indicates the electronic probability density function of the basis vectors. As shown in Fig. \ref{Rb(35)简并和非简并微扰}(b), the inset illustrates the typical electron probability density of the ``trilobite" molecules, the polar ULRMs, with black arrows pointing to the potential wells from which they originate. These molecules were first theoretically proposed and named by Greene et al.\cite{greene2000creation}. 

In the inset, the red and blue spheres represent the ground-state atom and the Rydberg atomic core, respectively. The distance between the two spheres is the internuclear distance, which is also the distance between the atomic core and the location of the maximum electronic probability density. This reflects the localization of the Rydberg electron near the ground-state atom (the perturber), which explaining why ``trilobite" molecules can exhibit PEDMs of several thousand Debye\cite{li2011homonuclear, booth2015production, althon2023exploring}. Additionally, an interesting phenomenon is seen: the number of lobes in the ``trilobite" molecule's wavefunction can be predicted. Starting from larger internuclear distances, the probability density of the ``trilobite" molecule corresponding to the first potential well is symmetric with two lobes, the second potential well corresponds to a symmetric four-lobe structure, and the \(x\)-th potential well corresponds to a symmetric \(2x\)-lobe pattern.

\begin{figure}[h]
\centering
\includegraphics[width=1\linewidth]{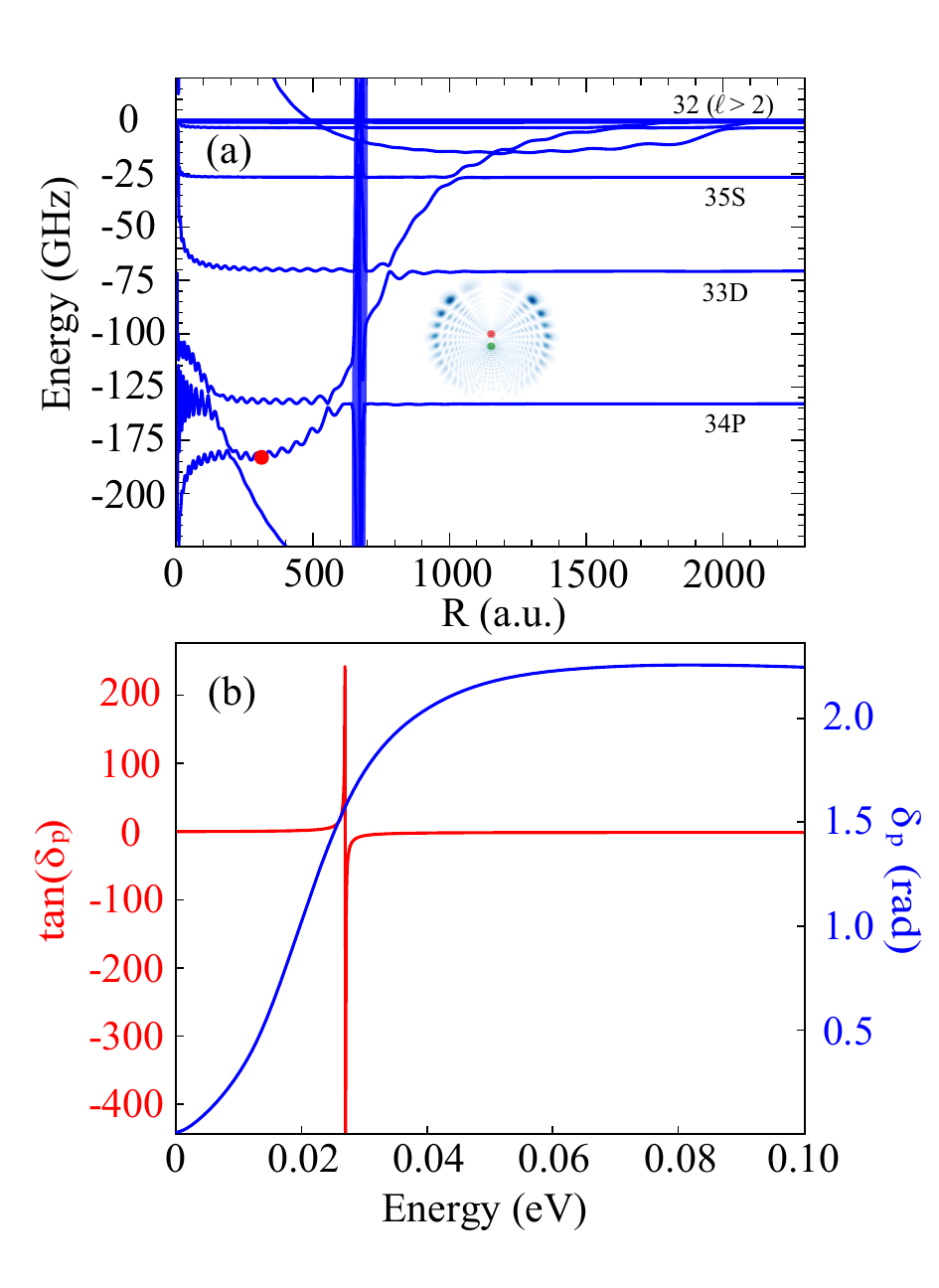}
\caption {Rb(\(n=35\)) ULRMs are formed by both \(s\)-wave scattering and \(p\)-wave interactions. (a) The blue solid line shows the PECs near the \(35S\) state, with the zero-energy reference at the \(n=32\) hydrogen atom line. In the region marked by the thick blue line around 680 a.u., \(p\)-wave scattering induces shape resonance, causing the previously isolated 32(\(l>2\)), \(35S\), \(33D\), and \(34P\) to couple, forming the ``butterfly" ULRMs at the red dot (around 308 a.u.). The inset shows the typical probability density plots of the ``butterfly" ULRMs, where the red (green) spheres represent the positions of the ground-state atom (atomic core). (b) The $^3S_1$ \(p\)-wave scattering phase shift for Rb is plotted. The blue line represents the relationship between the \(p\)-wave scattering phase shift and the electron energy, \(E = {k^2}/{2}\), for the ground-state Rb atom. The red line corresponds to the curve obtained by taking the tangent of the $^3S_1$ \(p\)-wave scattering phase shift. A resonance occurs around 0.026 eV, where \(\tan(\delta_p)\) diverges, corresponding to the thick blue line region in (a).
}
\label{P波形状共振}
\end{figure}

\subsection{Including \(P\)-wave interaction}

When considering the second term in Eq. (\ref{eq:V}), i.e., the \(p\)-wave scattering interaction, shape resonance occurs. The blue curve in panel (b) of Fig. \ref{P波形状共振} represents the $^3S_1$ \(p\)-wave scattering phase shift for Rb ground-state atoms scattering with Rydberg electron, and the red curve shows the tangent of this phase shift. It can be seen that when the electron energy \(E = {k^2}/{2}\) reaches 0.026 eV, the $^3S_1$ \(p\)-wave scattering phase shift approaches \(\pi/2\), and its tangent diverges. This phenomenon is referred to as shape resonance \cite{hamilton2002shape}. Therefore, \(a_p^3(k) = -{\tan(\delta_p(k))}/{k^3}\) also diverges at this point, resulting in the coupling of the originally well-isolated Rydberg states.

As a result, when the \(p\)-wave scattering interaction is included, the basis set used for calculating the high-\( l \) states must be expanded. We used all Rydberg states in Rb corresponding to the hydrogen levels \(n = 32\) and \(n = 31\), including \(32(l>2)\), \(35S\), \(34P\), \(33D\), and \(31(l>2)\), as the basis vectors for the calculation. The resulting PECs and typical electronic probability densities are shown in Fig. \ref{P波形状共振}(a), calculated in a manner similar to that for previous high-\( l \) states.

The divergence of the \(p\)-wave interaction near \(R = 680 \, \text{a.u.}\) corresponds to the divergence region of \(\tan(\delta_p)\) in Fig. \ref{P波形状共振}(b). Due to the nonphysical divergence, the PECs obtained near this region are unreliable and are covered by a thick blue line in Fig. \ref{P波形状共振}(a). At the left end of this blue line, the PECs of the \(34P\) state form a ``butterfly" molecule at the potential well around \(R = 308 \, \text{a.u.}\) (red dot), and its typical electronic probability density is shown in the inset. The shape of the molecule resembles a large butterfly, which is the origin of its name \cite{hamilton2002shape}.

\section{Heteronuclear ULRMs}
\subsection{Diatomic heteronuclear ULRMs}
\begin{figure}[h]
\centering
\includegraphics[width=1\linewidth]{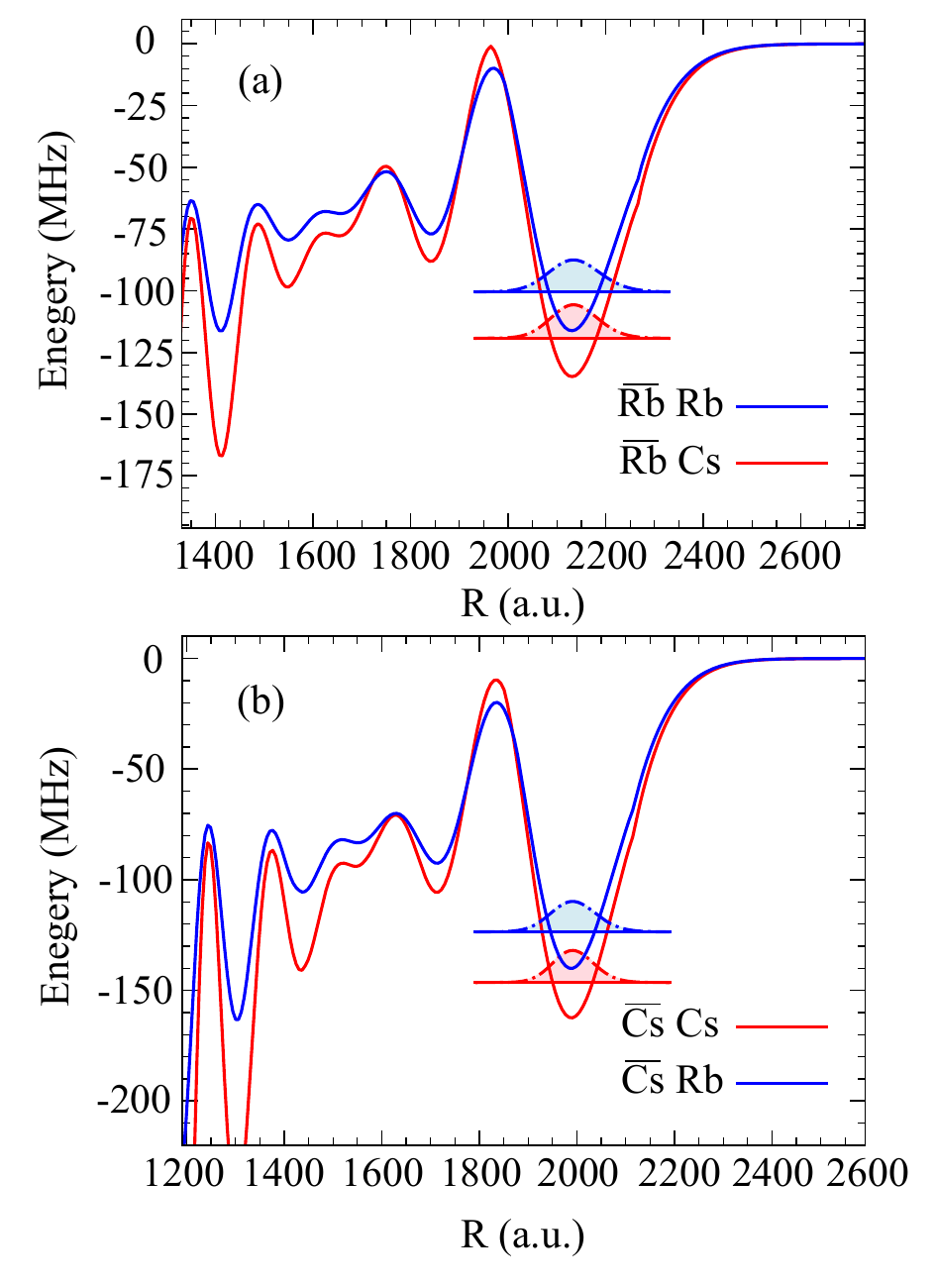}
\caption {PECs of ULRMs (\(35D\)) composed of Rb and Cs. (a) shows Rb as the Rydberg atom, and (b) shows Cs as the Rydberg atom, respectively. The zero-energy line of the curves corresponds to the energy of the isolated \(35D\) Rydberg atom. The red (blue) solid line illustrates the PECs of molecules including Rb(Cs) ground state atom. In the legend, atoms with a horizontal line above their symbols represent Rydberg atoms in ULRMs, while those without the line represent ground-state atoms. The vibrational ground-state energy level of the outer potential well is indicated by a solid horizontal line, and the vibrational ground-state wave function is represented by a dash-dot line. The area between these two lines is highlighted in light blue or light red.
}
\label{异核PECs}
\end{figure}

\begin{figure}[h]
\centering
\includegraphics[width=1\linewidth]{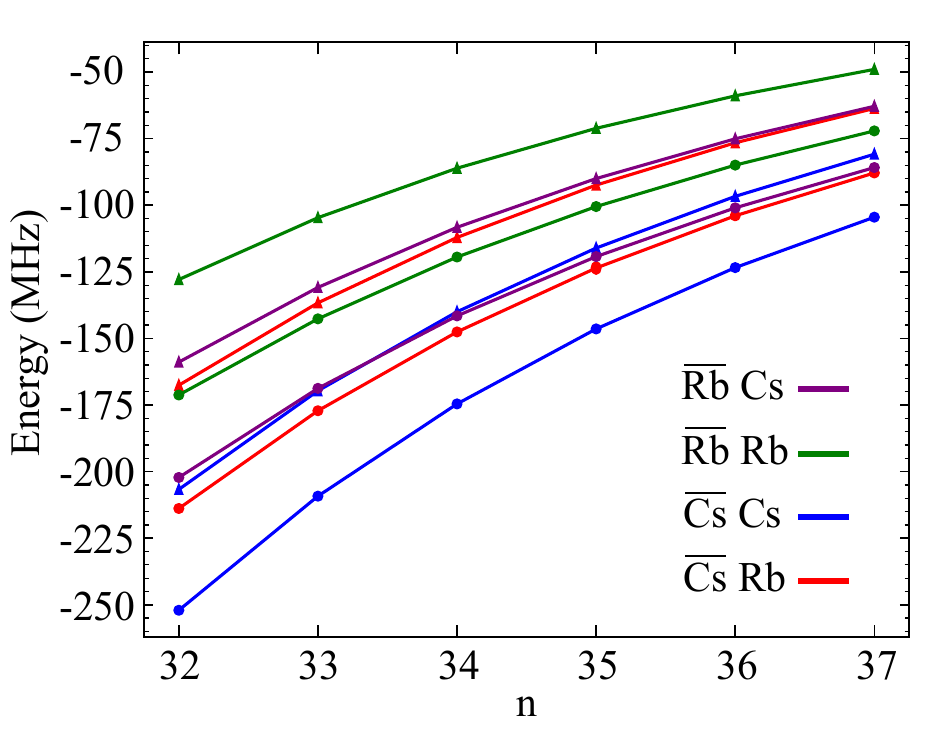}
\caption {Vibrational energy levels of ULRMs. The plot illustrates the vibrational ground-state and first excited-state energy levels of the outermost potential wells for \(D\)-state ULRMs with different \( n \). The curves formed by triangles (circles) represent the first excited-state (ground-state). The vertical axis represents the shift of the vibrational energy relative to the energy level associated with the isolated Rydberg atom. As indicated by the legend, the \(\overline{\text{Rb}}\text{ Rb}\), \(\overline{\text{Rb}}\text{ Cs}\), \(\overline{\text{Cs}}\text{ Cs}\), and \(\overline{\text{Cs}}\text{ Rb}\) lines represent the trends of the ground-state energy levels as functions of \( n \) for four different nuclear combinations.
}
\label{异核束缚能}
\end{figure}

\begin{figure}[!]
\centering
\includegraphics[width=\linewidth]{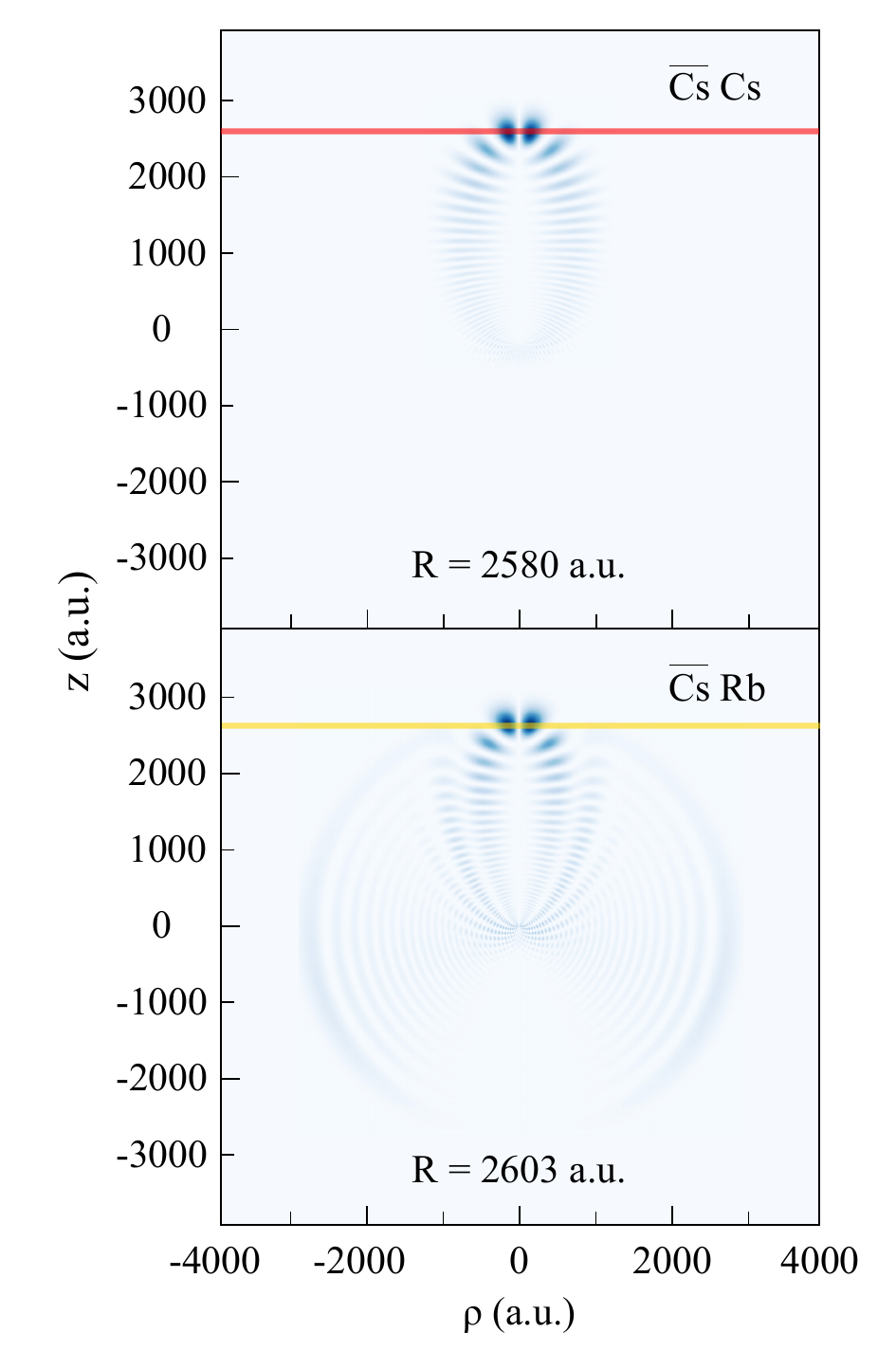}
\caption{Cylindrical coordinate surface plots of the typical electronic probability densities, \( \rho |\psi(\rho, z, 0)|^2 \) and \( \rho |\psi(\rho, z, \pi)|^2 \), for \(42S\) ULRMs of \( \overline{\text{Cs}}\)\text{ Cs} and \( \overline{\text{Cs}}\)\text{ Rb}. These plots correspond to the deepest point of the outermost potential well. The \( z \)-axis is aligned parallel to the line connecting the ground-state atom and the ionic nucleus, serving as the axis of rotational symmetry for the molecular electronic probability density. The ionic nucleus is fixed at the origin, while the ground-state atom is located at \( \rho = 0, z = R \), where \( R \) represents the internuclear distance. The positions \( R = 2580 \, \mathrm{a.u.} \) (red line) and \( R = 2603 \, \mathrm{a.u.} \) (yellow line) are indicated for the upper and lower plots. Darker shading represents regions of higher electronic probability density, indicating a greater likelihood of finding the electron. It can be seen that, due to the small non-integer quantum defect in the \(nS\) Cs atom, a significant portion of trilobite molecules mixes into the molecules, leading to electron localization near the ground-state atom. Moreover, compared to \( \overline{\text{Cs}}\)\text{ Cs} homonuclear ULRMs, the proportion of trilobites is smaller in \( \overline{\text{Cs}}\)\text{ Rb} heteronuclear ULRMs.
}
\label{异核电子概率密度}
\end{figure}

As shown in Eq. (\ref{eq:V}), the scattering length of the ground-state atom plays a pivotal role in determining the bond strength of the ULRMs, while the properties of the Rydberg atom govern the energy level structure of these molecules. Consequently, if the two atoms that form the ULRMs belong to different species, heteronuclear ULRMs with a broader range of physical properties can be realized. In recent years, heteronuclear ULRMs have been extensively studied both experimentally and theoretically \cite{whalen2020heteronuclear, eiles2018twocomponent, peper2021heteronuclear}. However, no detailed comparison has been made between the vibrational energy levels of the outermost potential wells of homonuclear and heteronuclear \(nD\) ULRMs. The PECs (\(35D\)) that include \(p\)-wave interactions, vibrational energy levels, and vibrational wave functions of heteronuclear Rydberg molecules are presented in Fig. \ref{异核PECs}. The following observations can be made:
1. Because of the larger absolute value of the zero-energy scattering length for Cs atoms, the outer potential well of the PECs is deeper when Cs serves as the ground-state atom compared to that of Rb.  
2. For a given type of Rydberg atom, the positions of the potential wells remain unchanged, indicating that the molecular size remains consistent.  

The vibrational ground-state and first excited-state binding energies of the four molecules as a function of \(n\) are shown in Fig. \ref{异核束缚能}. It is evident that the absolute value of the vibrational binding energy decreases with increasing \(n\), following the relationship \( E_{v0} \propto n^{-6} \). In particular, the four ground-state energy levels and the four first excited-state energy level curves do not intersect, demonstrating that the relative depths of the potential wells for these molecules remain unchanged over a wide range of the \(n\), which is a universal phenomenon. We predict that in ultracold Rb and ultracold Cs two-species atomic gases, when the excitation light frequency scans near the Rydberg state, these four types of ULRMs will form simultaneously. However, the formation probability is influenced by factors such as the potential well depth of the associated molecular states, interatomic distances, and the temperature and density of the Rb and Cs atomic ensembles.

We now focus on the heteronuclear molecular electronic probability density and PEDMs. Li et al. successfully produced \(nS\) Rb\(_2\) ULRMs using traditional two-photon association methods\cite{li2011homonuclear}. Due to the proximity of the Rb \(nS\) energy levels to the \((n-3)(l>2)\) Rb levels, the resulting \(nS\) Rb\(_2\) ULRMs were mixed with a small fraction of ``trilobite" molecules, leading to the appearance of PEDMs for the \(nS\) Rb\(_2\) ULRMs. This marked the first observation of PEDMs in homonuclear molecules \cite{li2011homonuclear}. Booth et al. later created \(nS\) Cs\(_2\) ULRMs in the laboratory. Due to the smaller non-integer quantum defect in the \(nS\) Cs state, which causes near-degeneracy with the \((n-4)(l>2)\) Cs energy levels, strong coupling between these states occurred. As a result, the \(nS\) Cs\(_2\) ULRMs were strongly mixed with ``trilobite" molecules, resulting in PEDMs on the order of kilo-Debye \cite{booth2015production}.

Given that the energy level structure of ULRMs is determined by the Rydberg atoms that constitute them, a similar phenomenon is expected to occur in \(\overline{\text{Cs}}\)\text{ Rb} ULRMs. Figure. \ref{异核电子概率密度} shows the typical electronic probability density at the deepest point of the outermost potential well of the \(42S\) state PECs. It can be seen that there is also a considerable admixture of trilobite molecules in \(42S\) \(\overline{\text{Cs}}\)\text{ Rb} ULRMs. However, compared to \(\overline{\text{Cs}}\)\text{ Cs}, the proportion of trilobite molecules in \(\overline{\text{Cs}}\)\text{ Rb} is lower, resulting in smaller PEDMs. The PEDMs of \(\overline{\text{Cs}}\)\text{ Rb} (\(\overline{\text{Cs}}\)\text{ Cs}) ULRMs are calculated to be 1670 (2081) Debye using $d(R) = \langle \Psi(R; \mathbf{r}) | z | \Psi(R; \mathbf{r}) \rangle$. The PEDMs of molecules can be measured by observing the broadening of spectra in a weak electric field, i.e., the Stark spectra\cite{althon2023exploring, booth2015production, li2011homonuclear}. Therefore, when an electric field is applied to ULRMs, the broadening will be smaller for \(\overline{\text{Cs}}\)\text{ Rb} compared to \(\overline{\text{Cs}}\)\text{ Cs}. These heteronuclear ULRMs have significant potential for applications in quantum information processing, quantum simulation, and precision measurements. Their unique electronic properties including large PEDMs and tunable interaction ranges make them ideal candidates for exploring novel phenomena in many-body physics and developing next-generation quantum technologies, such as probing spatial correlations in ultracold atomic gases\cite{manthey2015dynamically, whalen2019formation, whalen2019probing} and characterisation of low-energy electron-atom collisions\cite{anderson2014photoassociation, sassmannshausen2015experimental, bottcher2016observation, exner2024high}.

\subsection{Polyatomic heteronuclear ULRMs}

Normally, ULRMs consist of a Rydberg atom and a ground-state atom. However, Liu et al. theoretically investigated the formation and properties of polyatomic ULRMs and found that Ne atoms with positive scattering lengths can form trimers, even though the scattering interaction between Ne atoms and electrons is repulsive \cite{liu2006polyatomic, liu2009ultra}. Subsequently, Gaj et al. experimentally observed the spectra of polyatomic ULRMs\cite{gaj2014molecular}. They discovered that for an \(nS\) Rydberg Rb atom, it can form polyatomic ULRMs by simultaneously capturing multiple ground-state atoms. Due to the isotropy of the \(nS\) Rydberg atom wavefunction and the negligible interactions between the captured ground-state atoms, when the number of captured ground-state atoms is small, the binding energy of the molecule is an integer multiple of the dimer energy, which is clearly presented in the molecular spectra. When the number of captured ground-state atoms becomes very large, the influence of multiple ground-state atoms on the molecular spectrum results in an overall shift of the atomic spectral lines \cite{gaj2014molecular}. However, for a \(nD\) Rydberg atom, its anisotropy allows for the formation of angular trimers, whose energies are not necessarily integer multiples of the dimer energies \cite{fey2019effective}. Additionally, Luukko et al. proposed that due to perturbation-induced quantum scarring and electron density localization on randomly occurring atom clusters, polyatomic trilobite molecules can also form in dense cold atomic gases with densities as high as $10^{16}/cm^3$\cite{luukko2017polyatomic}. Fey et al. analyzed the geometrical configurations of triatomic trilobite molecules and developed a simple building principle that predicts the equilibrium configurations of triatomic molecules \cite{fey2019building}.

\begin{figure}[!]
\centering
\includegraphics[width=\linewidth]{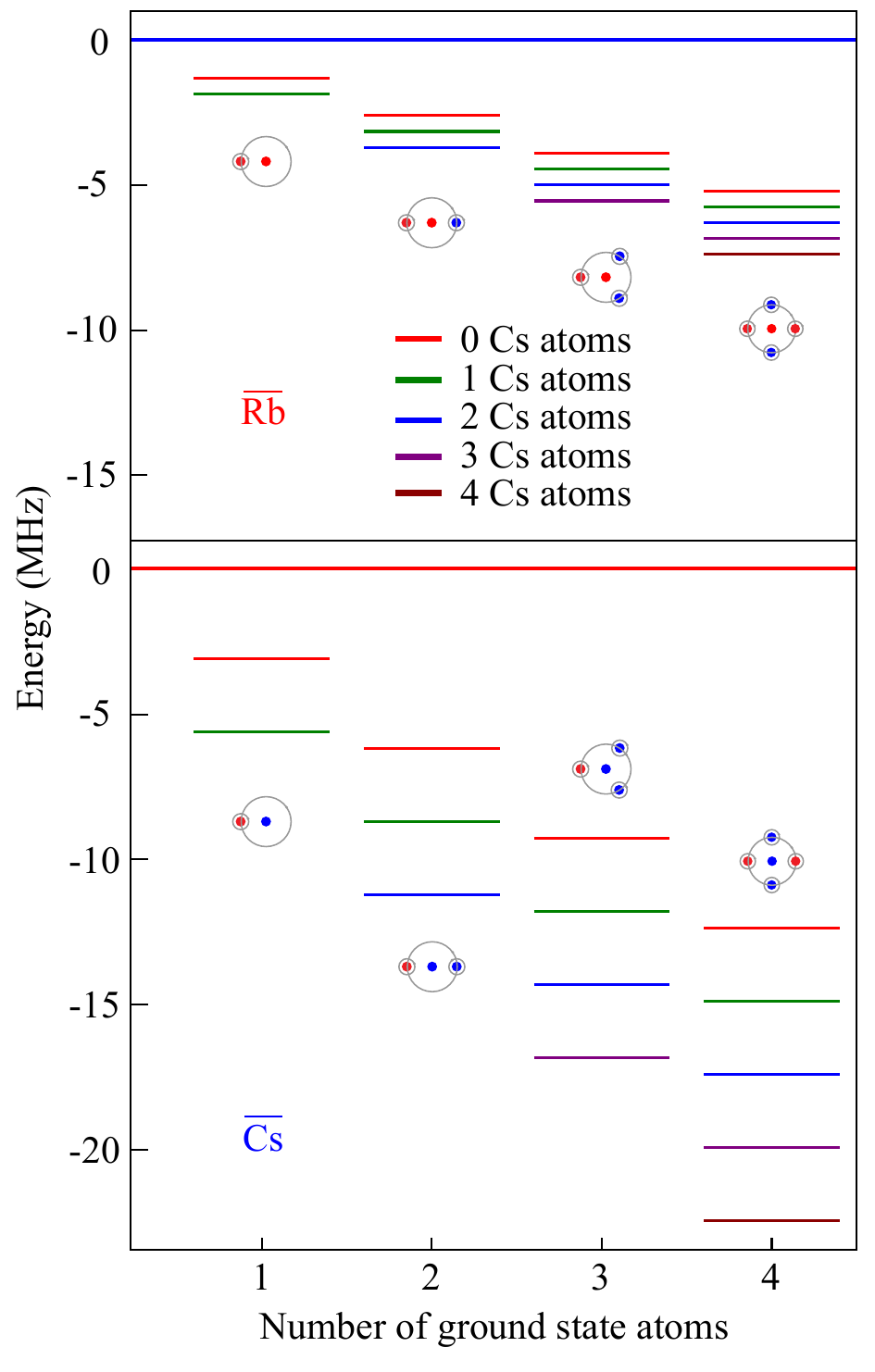}
\caption{Theoretical spectra of polyatomic heteronuclear ULRMs \((55S)\). The upper (lower) panel corresponds to Rb (Cs) atoms being excited to the Rydberg state. The blue and red lines at zero energy represents the energy level of an isolated Rydberg atom, specifically the Rb\((55S)\) level on the upper and the Cs\((55S)\) level on the lower. The spectra of various types of ULRMs are depicted, with red, green, blue, purple, and brown lines indicating the number of ground-state Cs atoms within the ULRMs as 0, 1, 2, 3, and 4, respectively. These lines reflect the changing binding energies as more ground-state Cs atoms are incorporated into the polyatomic ULRMs. Schematic representations of selected ULRMs are shown near the spectral lines, where red (blue) spheres denote Rb\textsuperscript{+} (Cs\textsuperscript{+}) ions, and gray regions illustrate the electrons and their orbitals. 
}
\label{多核异核分子理论预测谱}
\end{figure}

Building on these studies, we predict the existence of heteronuclear polyatomic ULRMs in ultracold Rb and Cs gases. In these systems, when either Rb or Cs atoms are excited to a Rydberg state, these Rydberg atoms can automatically capture multiple different ground-state atoms, forming polyatomic heteronuclear ULRMs. We discuss the heteronuclear polyatomic ULRMs composed of \(55S\) Rydberg atoms and predict their molecular spectra. We analyze how the binding energy and spectral features evolve as the number of ground-state atoms increases. In Fig. \ref{多核异核分子理论预测谱}, the upper (lower) panel shows the predicted spectra for heteronuclear ULRMs composed of \(55S\) Rb (Cs) Rydberg atoms. Since the intensity of the molecular vibrational spectra is proportional to the square of the bond length, and the intensity of the vibrational ground state is maximal, our theoretical predictions focus on the vibrational ground state energy levels in the outermost potential wells of the molecular PECs. Internal bound states formed by quantum reflection or ``butterfly" states arising from $p$-wave resonances are not considered.

For the \(55S\) state ULRMs, the outermost potential well vibrational ground state energies of the four types of dimers—$\overline{\text{Rb}}\text{ Rb}$, $\overline{\text{Rb}}\text{ Cs}$, $\overline{\text{Cs}}\text{ Rb}$, and $\overline{\text{Cs}}\text{ Cs}$—relative to their associated Rydberg levels are: $a = -1.3012\text { MHz}$, $b = -1.8470 \text{ MHz}$, $c = -3.0934 \text{ MHz}$, and $d = -5.6164 \text{ MHz}$. Following the approach of Gaj et al., if a polyatomic heteronuclear ULRM composed of a $55S$ Rb Rydberg atom contains $i$ Rb ground-state atoms and $j$ Cs ground-state atoms, the energy shift relative to the \(55S\) Rb Rydberg atom is given by $\Delta E_{\text{Rb}} = ia + jb$. Similarly, for polyatomic heteronuclear ULRMs composed of a Cs Rydberg atom, the energy shift is $\Delta E_{\text{Cs}} = ic + jd$. These calculated results are represented as energy level lines, as shown in Fig. \ref{多核异核分子理论预测谱}. They may serve as a guide for experimental observations and future applications in quantum technologies and molecular spectroscopy.

\section{Summary}

This paper presents a theoretical study of heteronuclear ULRMs, focusing on systems where Rb and Cs atoms form heteronuclear ULRMs. The study highlights the pivotal role of scattering length of ground-state atoms and the properties of Rydberg atoms in determining the bond strength and energy level structure of these molecules. The \(p\)-wave scattering interaction, particularly near shape resonance, plays a critical role in coupling Rydberg states, modifying potential energy curves (PECs) and molecular structures. The paper compares the vibrational energy levels and the influence of different atomic species on the energy level structure, focusing on their evolution with respect to \(n\). Additionally, the formation of polyatomic heteronuclear ULRMs is explored, including the impact of multiple ground-state atoms on the molecular spectra. The paper predicts the energy shifts for different configurations of Rb and Cs atoms in polyatomic ULRMs and provides theoretical spectra that could guide experimental investigations. These findings pave the way for future experimental studies of heteronuclear ULRMs and their applications, such as study of pair-correlation functions in atomic mixtures and the dynamics of complex many-body heteronuclear systems. 

\section{ACKNOWLEDGMENTS}

We acknowledge Weibin Li for providing the computational codes used to calculate the PECs with only s-wave interactions. We acknowledge funding from the National Key R\&D Program of China (Grant No. 2022YFA1404002), the National Natural Science Foundation of China (Grants No. T2495253, No. 61525504, and No. 61435011), the Major Science and Technology Projects in Anhui Province (Grant No. 202203a13010001).

\section{Author contribution}
D-S. D. formulated the plan and provided support for this project. Q. L. performed the theoretical calculations and wrote the manuscript, and S-Y. S. provided many valuable insights. All authors participated in the discussions and provided valuable suggestions.

\FloatBarrier
\bibliography{ref}
\end{document}